# Dyadic Green's function for the graphene-dielectric stack with arbitrary field and source points

Shiva Hayati Raad,[1] Zahra Atlasbaf,[1,*] Mauro Cuevas[2,3]

[1] *Department of Electrical and Computer Engineering, Tarbiat Modares University, Tehran, Iran*
[2] *Consejo Nacional de Investigaciones Científicas y Técnicas (CONICET), Buenos Aires, Argentina*
[3] *Facultad de Ingeniería, Universidad Austral, Mariano Acosta 1611, Pilar, Buenos Aires, Argentina*

*Corresponding author: atlasbaf@modares.ac.ir*

**In this paper, dyadic Green's function for a graphene-dielectric stack is formulated based on the scattering superposition method. To this end, scattering Green's function in each layer is expanded in terms of cylindrical vector wave functions with unknown coefficients. Using the Kronecker delta function in the field expansion, it is considered that the field and source points lie in the arbitrary layers. Afterward, recurrence relations for calculating the unknown expansion coefficients are derived by applying the impedance boundary conditions at the interface of a graphene sheet surrounded by two adjacent dielectric layers. The verification of the calculated coefficients is conducted by utilizing them in the analysis of graphene-based structures with different numbers of layers, including 1) free-standing frequency-selective surfaces (FSSs) and 2) parallel plates (PPs) with graphene walls. A potential application of our proposed structure is investigating the interaction of donor-acceptor pairs resided in the arbitrary layers of the graphene-dielectric stack with a desired number of layers.**

http://dx.doi.org/10.1364/AO.99.099999

The integral equation (IE) based numerical techniques are among the most efficient approaches for the rigorous analysis of the planarly stratified media [1]. Hence, Green's function derivation for these structures is of great interest. Green's function is commonly represented in terms of highly oscillatory and slowly decaying Sommerfeld's integrals (SIs) [2] and generalized pencil of function method (GPOF) has been proposed for the efficient numerical calculation of SIs [3]. In general, Green's function technique can be successfully used in the analysis of structures formed by different types of materials and for various configurations, including but not limited to surface optics [4], layered chiral media [5], inhomogeneous optical media [6], anisotropic media [7, 8], eccentrically stratified sphere [9], a cluster of spheres [10], and perfect electromagnetic conductor (PEMC) cylinder [11].

Recently, graphene material has been widely used in the design of optical devices. For the analysis of graphene-based devices, its surface conductivity has been incorporated in various numerical techniques such as the method of moment (MoM) [12], finite-difference time-domain (FDTD) [13], finite element method (FEM) [14], and circuit theory [15]. Dyadic Green's function (DGF) for a graphene-based stratified planar medium has not been reported yet and will be derived in this research. To this end, eigenfunction expansion in the framework of the scattering superposition method will be employed. The proposed formulation provides time and memory-efficient exact recurrence formulas for the analysis of graphene-based stacks with the desired number of layers. A similar method has been used to extract the DGF of graphene-based multi-layered spherical structures [16].

The extracted Green's function can be possibly used in the analysis and design of different optical devices and applications. For instance, the interaction of quantum emitters with the surrounding media can be analyzed using the associated Green's function because of the feasibility of approximating the emitters with electric dipoles in the physical models [17]. Moreover, it is demonstrated that graphene-based parallel-plate waveguides and hyperbolic metamaterials exhibit a giant Purcell effect, beneficial for enhancing the spontaneous emission of the dipole emitter [18, 19]. Also, multi-periodic hyperbolic like metamaterials have been proposed as another method for the enhancement of the Purcell effect [20]. Therefore, our proposed approach may be valuable for developing other novel substrates for spontaneous emission enhancement by engineering the optical, material, and geometrical parameters. The paper is organized as follows. After introducing the cylindrical vector wave functions in Section 2, they are used in the electromagnetic field expansion. Later, the unknown expansion coefficients are obtained by applying the boundary conditions, considering the surface currents on graphene interfaces. In Section 3, the validity of the extracted formulas is verified by analyzing graphene-based structures with different numbers of layers. For simplicity, two special cases of source-free media and plane wave illumination are considered.

## 1. Dyadic Green's function formulation

In this section, the DGF of an N-layer graphene-dielectric stack, as shown in Fig. 1, will be formulated via the scattering superposition method. This geometry can be used to design various optical devices by properly choosing the optical, geometrical, and material properties of the layers. To provide a powerful tool for the analysis of these devices, an impulse source is considered in arbitrary layer $j$ (source point) and the resulted fields are calculated in another arbitrary layer $i$ (field point).

The surface conductivities of graphene interfaces are calculated based on the Kubo formulas, where chemical potential $\mu_c$ and relaxation time $\tau$ influence their optical properties [21]. Incorporating the graphene surface conductivity in the spectral and time-domain Green's function of the planar medium is of interest and has been done for the single graphene sheet and double-stacked graphene sheets previously [18, 22-25]. It is worth noting that by modeling each graphene sheet as a thin dielectric [26], the corresponding Green's function can be attained by the Green's function of planarly stratified media [27]. One advantage of the former approach is that each graphene interface is treated with the two-sided boundary condition and it does not add a further 3D layer to the structure. This feature is of great importance in many-layer structures. Moreover, the solution of the electromagnetic problems with the impedance boundary condition has been extensively studied in the literature [28-32]. By considering the surface impedance as the inverse of the graphene surface conductivity, the research can be used to consider other types of interfaces constructed by two-dimensional materials.

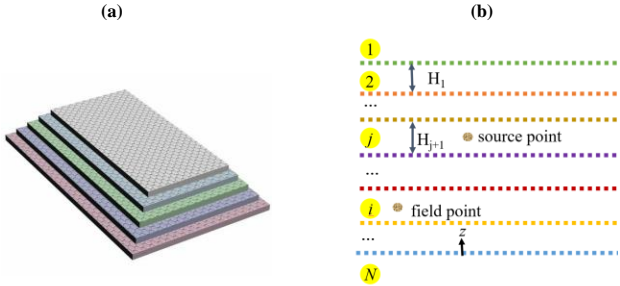

Fig 1. Multi-layered planar structure with graphene interfaces, (a) 3D and (b) 2D views. Dashed lines at the interfaces in the 2D view represent graphene sheets. This geometry can be used to design various optical devices by properly choosing the optical, geometrical, and material properties of the layers. To provide a powerful tool for the analysis of these devices, an impulse source is considered in arbitrary layer $j$ and the resulted fields are calculated in arbitrary layer $i$. The constitutive parameters of arbitrary layer i are denoted by ($\varepsilon$, $\mu$) during the formulation.

### A. Vector wave functions in the cylindrical coordinate

The first step in formulating planar structures is to choose the coordinate system since for these geometries both rectangular and cylindrical systems are applicable [33, 34]. In the dyadic Green's function of the graphene sheet in the rectangular coordinate, double integrals appear [35]. These integrals can be calculated with numerical methods such as the Romberg integration routine [24]. By converting the integrals to the polar coordinate, the angular integrals can be performed analytically [24]. Importantly, the DGF of the planar structures can be greatly simplified in the cylindrical coordinate system [23]. Therefore, the formulation is conducted in the cylindrical coordinate. Once the coordinate is chosen, the corresponding vector wave functions should be used in the expansion of the electromagnetic fields. A complete description of the cylindrical vector wave functions $M_{n\lambda}$ and $N_{n\lambda}$ are defined in [33]. The scalar wave function in the cylindrical coordinate for a layer with index $j$ is defined as:

$$\psi_\lambda = Z_n(\lambda r) e^{in\phi} e^{ih_j z} \tag{1}$$

where $Z_n$ is the Bessel function of the first kind or Hankel function of the first kind, both with the orders $n$. For the planarly layered media, the Bessel functions are used as the eigenfunction [27]. Moreover, parameters $\lambda$, $n$, and $h_j$ are wavenumbers in radial, azimuthal, and longitude directions. They are dependent via the wave number $k_j$ of the region as $\lambda^2 + h_j^2 = k_j^2$. It is essential to note that to account for the azimuthal variation in the description of the vector wave functions, exponential function rather than trigonometric functions are used. This assumption obviates the requirement for dealing with even and odd vector functions generated respectively by $\cos\phi$ and $\sin\phi$ dependencies [36]. Using the above $\psi_\lambda$, vector wave functions representing the electric field of transverse electric (TE) and transverse magnetic (TM) waves can be obtained respectively as [33, 36]:

$$M_{n\lambda}(h_j) = e^{in\phi} e^{ih_j z} \left[ in \frac{J_n(\lambda r)}{r} \hat{r} - \frac{\partial J_n(\lambda r)}{\partial r} \hat{\phi} \right] \tag{2}$$

$$N_{n\lambda}(h_j) = \frac{1}{k_j} e^{in\phi} e^{ih_j z} \times$$
$$\left[ ih_j \frac{\partial J_n(\lambda r)}{\partial r} \hat{r} - \frac{h_j n}{r} J_n(\lambda r) \hat{\phi} - \lambda^2 J_n(\lambda r) \hat{z} \right] \tag{3}$$

Note that the non-italic N letter represents the number of layers while the boldface italic $N$ letter denotes the vector wave function for TM waves. Also, when $i$ is subscript or superscript, it denotes the number of layers otherwise it denotes $i = \sqrt{-1}$.

### B. Electromagnetic field expansion

Once the vector wave functions are introduced, they can be used to expand the electromagnetic fields of each layer by assigning unknown coefficients to them. Due to the scattering superposition method, the electric field at layer $i$ resulting from a source at layer $j$, denoted by $\bar{G}_e^{(ij)}$, equals the sum of free-space Green's function $\bar{G}_{0e}$ and scattering Green's function $\bar{G}_{es}^{(ij)}$. On the other hand, $\bar{G}_{0e}$ is the field due to a source radiating in an unbounded medium and it can be found via $\bar{G}_m$ method. The $\bar{G}_m$ method can be described as the use of the Ohm-Rayleigh technique to extract the magnetic Green's function and later find the electric Green's function using Maxwell's equations. Due to the solenoidal nature of magnetic Green's functions, its eigenfunction expansion is simpler [33]. When the field point $\bar{R}$ and source point $\bar{R}'$ coincide, the free-space Green's function has a singularity in the form of $\hat{z}\hat{z}\delta(\bar{R}-\bar{R}')/k_j^2$, where $k_j$ denotes the wavenumber in the source medium [33]. The influence of dielectric layers and graphene interfaces are included in Green's function via $\bar{G}_{es}^{(ij)}$. Due to the isotropic nature of graphene surface conductivity, TE and TM waves are decoupled and one may write [27]:

$$\bar{G}_{es}^{(ij)}(\bar{R},\bar{R}') = \frac{i}{4\pi}\int_0^\infty d\lambda \sum_{n=-\infty}^\infty \frac{1}{\lambda h_j} \times$$
$$\{(1-\delta_i^N)M_{n\lambda}(h_i)\times[(1-\delta_j^1)A_M^{ij}M'_{n\lambda}(-h_j)+(1-\delta_j^N)B_M^{ij}M'_{n\lambda}(h_j)] +$$
$$(1-\delta_i^N)N_{n\lambda}(h_i)[(1-\delta_j^1)A_N^{ij}N'_{n\lambda}(-h_j)+(1-\delta_j^N)B_N^{ij}N'_{n\lambda}(h_j)] +$$
$$(1-\delta_i^1)M_{n\lambda}(-h_i)[(1-\delta_j^1)C_M^{ij}M'_{n\lambda}(-h_j)+(1-\delta_j^N)D_M^{ij}M'_{n\lambda}(h_j)] +$$
$$(1-\delta_i^1)N_{n\lambda}(-h_i)\times[(1-\delta_j^1)C_N^{ij}N'_{n\lambda}(-h_j)+(1-\delta_j^N)D_N^{ij}N'_{n\lambda}(h_j)]\}$$
(4)

where $A_{M,N}^{ij}$, $B_{M,N}^{ij}$, $C_{M,N}^{ij}$ and $D_{M,N}^{ij}$ are the unknown coefficients to be found by applying boundary conditions at interfaces. Also, $\delta_j^i$ represent the Kronecker delta function having a non-zero value for $i=j$. When calculating the interaction of donor-acceptor dipole emitters, the location of the emitters influences their optical responses [37]. Therefore, using the delta function, the position of the source and field points are considered to be arbitrary. Care must be taken to the summation index, where it is varying from minus infinity to infinity due to the exponential dependency of the azimuthal function. Moreover, the Sommerfeld integral of (4) has no closed-form analytical solution, but in the far-zone, it can be approximated by the saddle-point method [33].

Finally, the unknown expansion coefficients are determined by applying the boundary conditions at the interface of two adjacent layers. The linear system of equations resulting from the tangential components of the electric field can be written as Eq. 6(a) of [27], extracted for dielectric interfaces. Hence:

$$A_M^{ij}e^{ih_iz_i} + C_M^{ij}e^{-ih_iz_j} = (A_M^{(i+1)j}+\delta_{i+1}^j)e^{ih_{i+1}z_i} + C_M^{(i+1)j}e^{-ih_{i+1}z_i} \quad (5)$$

$$B_M^{ij}e^{ih_iz_i} + (D_M^{ij}+\delta_j^i)e^{-ih_iz_j} = B_M^{(i+1)j}e^{ih_{i+1}z_i} + D_M^{(i+1)j}e^{-ih_{i+1}z_i} \quad (6)$$

$$\frac{h_i}{k_i}\left[A_N^{ij}e^{ih_iz_i}+C_N^{ij}e^{-ih_iz_j}\right] = \frac{h_{i+1}}{k_{i+1}}\left[(A_N^{(i+1)j}+\delta_{i+1}^j)e^{ih_{i+1}z_i}+C_N^{(i+1)j}e^{-ih_{i+1}z_i}\right]$$
(7)

$$\frac{h_i}{k_i}\left[B_N^{ij}e^{ih_iz_i}+(D_N^{ij}+\delta_j^i)^{-ih_iz_j}\right] = \frac{h_{i+1}}{k_{i+1}}\left[B_N^{(i+1)j}e^{ih_{i+1}z_i}+D_N^{(i+1)j}e^{-ih_{i+1}z_i}\right]$$
(8)

For the tangential components of the magnetic field, surface currents due to the presence of graphene should be considered. Therefore:

$$\hat{z}\times\left(\frac{\nabla\times\bar{G}_e^{[(i+1)j]}}{\mu_{i+1}}-\frac{\nabla\times\bar{G}_e^{(ij)}}{\mu_i}\right) = -i\omega\sigma_{(i+1)i}\hat{z}\times\hat{z}\times\bar{G}_e^{(ij)} \quad (9)$$

where $\sigma_{(i+1)i}$ is the scalar surface conductivity of graphene at the interface of $i+1$ and $i$ layers. The linear system of equations obtained by applying the boundary condition (9) to the field expansion in (4) equals:

$$\frac{h_{i+1}}{\mu_{i+1}}\left[(A_M^{(i+1)j}+\delta_{i+1}^j)e^{ih_{i+1}z_i}-C_M^{(i+1)j}e^{-ih_{i+1}z_i}\right] -$$
$$\frac{h_i}{\mu_i}\left[A_M^{ij}e^{ih_iz_i}+C_M^{ij}e^{-ih_iz_i}\right] = i\omega\sigma_{(i+1)i}\left[A_M^{ij}e^{ih_iz_i}+C_M^{ij}e^{-ih_iz_i}\right]$$
(10)

$$\frac{h_{i+1}}{\mu_{i+1}}\left[B_M^{(i+1)j}e^{ih_{i+1}z_i}-D_M^{(i+1)j}e^{-ih_{i+1}z_i}\right] - \frac{h_i}{\mu_i}\left[B_M^{ij}e^{ih_iz_i}+(D_M^{ij}+\delta_j^i)e^{-ih_iz_i}\right]$$
$$= i\omega\sigma_{(i+1)i}\left[B_M^{ij}e^{ih_iz_i}+(D_M^{ij}+\delta_j^i)e^{-ih_iz_i}\right]$$
(11)

$$\frac{k_{i+1}}{\mu_{i+1}}\left[(A_N^{(i+1)j}+\delta_{i+1}^j)e^{ih_{i+1}z_i}-C_N^{(i+1)j}e^{-ih_{i+1}z_i}\right] -$$
$$\frac{k_{i+1}}{\mu_{i+1}}A\left[_N^{ij}e^{ih_iz_i}+C_N^{ij}e^{-ih_iz_i}\right] = \frac{h_i}{k_i}i\omega\sigma_{(i+1)i}\left[A_N^{ij}e^{ih_iz_i}+C_N^{ij}e^{-ih_iz_i}\right]$$
(12)

$$\frac{k_{i+1}}{\mu_{i+1}}\left[B_N^{(i+1)j}e^{ih_{i+1}z_i}-D_N^{(i+1)j}e^{-ih_{i+1}z_i}\right] - \frac{k_{i+1}}{\mu_{i+1}}\left[B_N^{ij}e^{ih_iz_i}+(D_N^{ij}+\delta_j^i)e^{-ih_iz_i}\right]$$
$$= \frac{h_i}{k_i}i\omega\sigma_{(i+1)i}\left[B_N^{ij}e^{ih_iz_i}+(D_N^{ij}+\delta_j^i)e^{-ih_iz_i}\right]$$
(13)

Pairing (5)-(8) respectively with (10)-(13), the coefficients of the $(i+1)$ layer can be written in terms of those of $i$ layer. Therefore, by defining equivalent reflection and transmission coefficients as in Table 1, one may obtain:

$$\begin{bmatrix} A_{M,N}^{(i+1)j}+\delta_{i+1}^j & B_{M,N}^{(i+1)j} \\ C_{M,N}^{(i+1)j} & D_{M,N}^{(i+1)j} \end{bmatrix} =$$
$$\begin{bmatrix} \frac{1}{T_{Fi}^{H,V}}e^{i(h_{i+1}-h_i)z_i} & \frac{R_{Fi}^{H,V}}{T_{Fi}^{H,V}}e^{-i(h_{i+1}+h_i)z_i} \\ \frac{R_{Pi}^{H,V}}{T_{Pi}^{H,V}}e^{i(h_{i+1}+h_i)z_i} & \frac{1}{T_{Pi}^{H,V}}e^{-i(h_{i+1}-h_i)z_i} \end{bmatrix}\begin{bmatrix} A_{M,N}^{ij} & B_{M,N}^{ij} \\ C_{M,N}^{ij} & D_{M,N}^{ij}+\delta_j^i \end{bmatrix}$$
(14)

where the first term on the right-hand side is denoted by $T_{M,N}^i$. Also, $H$ and $V$ represent the contribution of TE and TM waves, respectively. The subscript $F$ denotes the outgoing waves and the subscript $P$ is used for the incoming ones and $g = i\omega\mu_i\mu_{i+1}\sigma_{(i+1)i}$. The unknown coefficients can be calculated via the recurrence relations of [27] using $T_{M,N}^i$ matrix, where the extracted equivalent reflection and transmission coefficients must be used. For the sake of completeness, it should be noted that the unknown coefficients are calculated with the aid of $\bar{T}^{(K)}$ matrix that is constructed via the multiplication of successive $T^i_{M,N}$ matrices with the superscripts ranging from $i=$ N−1 to $K$.

**Table 1 Equivalent reflection and transmission coefficients for TE/TM waves**

| TE waves | |
|---|---|
| $R^H_{Fi} = \dfrac{\mu_i h_{i+1} - \mu_{i+1} h_i + g}{\mu_i h_{i+1} + \mu_{i+1} h_i + g}$ | (15) |
| $T^H_{Fi} = \dfrac{2\mu_i h_{i+1}}{\mu_i h_{i+1} + \mu_{i+1} h_i + g}$ | (16) |
| $R^H_{Pi} = \dfrac{\mu_i h_{i+1} - \mu_{i+1} h_i - g}{\mu_i h_{i+1} + \mu_{i+1} h_i - g}$ | (17) |
| $T^H_{Pi} = \dfrac{2\mu_i h_{i+1}}{\mu_i h_{i+1} + \mu_{i+1} h_i - g}$ | (18) |
| TM waves | |
| $R^V_{Fi} = \dfrac{\mu_i h_i k_{i+1}^2 - \mu_{i+1} h_{i+1} k_i^2 + g h_i h_{i+1}}{\mu_i h_i k_{i+1}^2 + \mu_{i+1} h_{i+1} k_i^2 + g h_i h_{i+1}}$ | (19) |
| $T^V_{Fi} = \dfrac{2\mu_i h_{i+1} k_i k_{i+1}}{\mu_i h_i k_{i+1}^2 + \mu_{i+1} h_{i+1} k_i^2 + g h_i h_{i+1}}$ | (20) |
| $R^V_{Pi} = \dfrac{\mu_i h_i k_{i+1}^2 - \mu_{i+1} h_{i+1} k_i^2 - g h_i h_{i+1}}{\mu_i h_i k_{i+1}^2 + \mu_{i+1} h_{i+1} k_i^2 - g h_i h_{i+1}}$ | (21) |
| $T^V_{Pi} = \dfrac{2\mu_{i+1} h_{i+1} k_i k_{i+1}}{\mu_i h_i k_{i+1}^2 + \mu_{i+1} h_{i+1} k_i^2 - g h_i h_{i+1}}$ | (22) |

## 2. Verification of the technique

In this section, the validity of the extracted formulas will be shown by analyzing two structures including 1) graphene sheet/nano patch at the interface of semi-infinite half-spaces [Fig. 2(a)] 2) two parallel graphene plates/nano-patches [Fig. 2(b)]. The assumption of nano-patch is to improve the graphene light-matter interaction [38]. The surface impedance of electrically small graphene patches can be obtained as [39, 40]:

$$Z_p = \left(0.6\left[D_0/(D_0-g)\right]^3 + 0.4\right)Z_g + \frac{i}{2}\left(-\frac{k_0 D_0 \sqrt{\varepsilon_{eff}}}{\pi}\ln\left[\sin\left(\frac{\pi g}{2D_0}\right)\right]\right)^{-1}\sqrt{\frac{\mu_0}{\varepsilon_0 \varepsilon_{eff}}} \quad (23)$$

where the graphene surface impedance $Z_g$ is defined as the inverse of the graphene surface conductivity $\sigma$. Also, $D_0$, $k_0$, and $g$ are the periodicity, free-space wavenumber, and gap size, respectively.

### A. Use of DGF for radiating and guiding problems
To avoid the solution of the SIs during the verification procedure, the point source is considered in the region (1) and it is moved to infinity to produce a plane wave. The reflectance (in layer 1) and transmittance (in layer N) of a plane wave can be calculated respectively using $B^{11}_{M,N}$ and $D^{N1}_{M,N}$ coefficients defined as [27, 33]:

$$B^{11}_{M,N} = -\left[\frac{T^{(1)}_{12}}{T^{(1)}_{11}}\right]_{M,N} \quad (24)$$

$$D^{N1}_{M,N} = \left[-\frac{T^{(1)}_{21}T^{(1)}_{12}}{T^{(1)}_{11}} + T^{(1)}_{22}\right]_{M,N} \quad (25)$$

For normally incident plane wave: $h_i = k_i$. It should be noted that prior to this formulation, the scattering matrix method has been used to derive the amplitude coefficients of the suspended graphene stacks under TM wave illumination [41]. Although this formulation has resulted in similar coefficients as our proposed dyadic technique, our formulation is more general since it treats TE and TM waves simultaneously. Moreover, we have generalized the method by considering a point dipole as the source which results in including more expansion coefficients. In the scattering matrix method, a plane wave is considered as the illuminating source. In comparison to the T-matrix method, the presented formulation which is called the boundary condition method is more simple when many electric and magnetic sources exist in the medium [27]. Finally, by calculating the poles of Green's function expansion coefficients, the propagating and attenuation constants of the waveguides produced by planar graphene-based structures can be achieved.

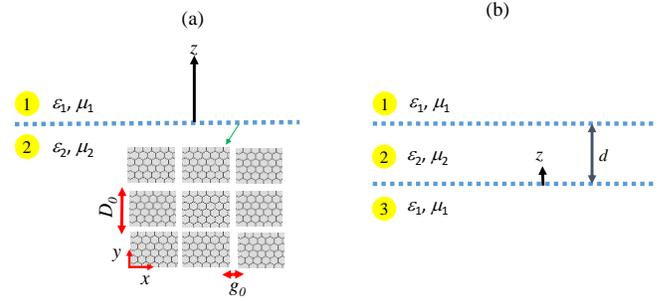

Fig 2. (a) Free-standing graphene sheet/nano-patch as a frequency-selective surface and (b) two stacked graphene plates/nano-patches. The consideration of the nano-patch is for the improvement of the light-matter interaction. In these structures, the plane wave reflection and transmission coefficients are calculated and compared with those obtained by the TMM method. For this purpose, the point dipole is considered in region 1 and is moved to infinity. The polarization of the plane wave is recognized by the type of the vector wave function (**M** denoting TE waves and **N** denoting TM waves).

### B. Free-standing graphene sheet/nano-patch

The first example is a free-standing graphene sheet as shown in Fig. 2(a). For the double-layered structure: $T^{(1)}_{M,N} = T^1_{M,N}\big|_{z_1=0}$. Under plane wave illumination the non-zero expansion coefficients read as:

$$B^{11}_{M,N} = -R^{H,V}_{F1} \quad (26)$$

$$D^{21}_{M,N} = \frac{1}{T^{H,V}_{P1}}\left(1 - R^{H,V}_{P1} R^{H,V}_{F1}\right) \quad (27)$$

After some algebraic manipulations, the above coefficients are simplified and summarized in Table 2, where $\sigma=\sigma_{21}$. These equations can be found in [22], extracted based on the Hertzian potentials.

**Table 2 Reflection and transmission coefficients of a normally incident plane wave with TE and TM polarizations illuminating to a graphene sheet/nano-patch**

| | |
|---|---|
| $B_M^{11} = \dfrac{\mu_2 k_1 - \mu_1 k_2 - i\omega\sigma\mu_1\mu_2}{\mu_2 k_1 + \mu_1 k_2 + i\omega\sigma\mu_1\mu_2}$ | (28) |
| $B_N^{11} = \dfrac{\varepsilon_2 k_1 - \varepsilon_1 k_2 - i\sigma k_1 k_2 / \omega}{\varepsilon_2 k_1 + \varepsilon_1 k_2 + i\sigma k_1 k_2 / \omega}$ | (29) |
| $D_M^{21} = \dfrac{2\mu_2 k_1}{\mu_2 k_1 + \mu_1 k_2 + i\omega\sigma\mu_1\mu_2}$ | (30) |
| $D_N^{21} = \dfrac{2\varepsilon_2 k_1}{\varepsilon_2 k_1 + \varepsilon_1 k_2 + i\sigma k_1 k_2 / \omega}$ | (31) |

The graphical representation of the reflection (R) and transmission (T) coefficients for the free-standing period patches of Fig. 2(a) with $D_0=5$ μm, $g_0=0.5$ μm, $\mu_c=0.5$ eV, and $\tau=0.5$ ps are illustrated in Fig. 3. The results are in good agreement with those regenerated with the transfer matrix method (TMM) [42]. The TMM approach is generally used to calculate the electromagnetic response of the plane wave or solution of source-free Maxwell's equations [19, 41, 43, 44]. Moreover, as an alternative method, Green's function coefficients can be extracted by the transfer matrix method [45]. Here, the investigation of plane wave scattering analysis using TMM method is of interest.

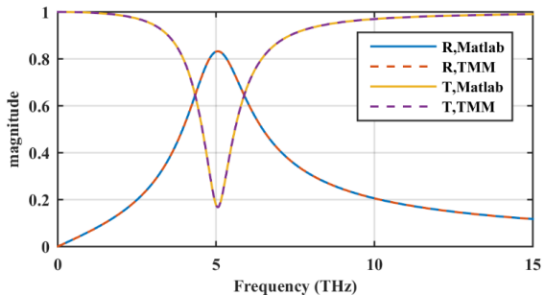

Fig 3. The reflection ($R$) and transmission ($T$) coefficients of a plane wave incident on the free-standing period patches of Fig. 2(a) with $D_0=5$ μm, $g_0=0.5$ μm, $\mu_c=0.5$ eV, and $\tau=0.5$ ps.

### C. Parallel graphene plates/nano-patches

The second example is a parallel plate waveguide (PPWG) with electrically biased graphene walls as shown in Fig. 2(b). The characteristic equation for the surface waves has been obtained by the transverse resonance method [46]. To derive the characteristic equation through DGF, by calculating the $T_M^{(1)}$ matrix of TE waves as:

$$T_M^{(1)} = T_M^2 \cdot T_M^1 =$$
$$\begin{bmatrix} \dfrac{1}{T_{F2}^H} e^{i(h_2-h_1)d} & \dfrac{R_{F2}^H}{T_{F2}^H} e^{-i(h_2+h_1)d} \\ \dfrac{R_{P2}^H}{T_{P2}^H} e^{i(h_2+h_1)d} & \dfrac{1}{T_{P2}^H} e^{-i(h_2-h_1)d} \end{bmatrix} \cdot \begin{bmatrix} \dfrac{1}{T_{F1}^H} & \dfrac{R_{F1}^H}{T_{F1}^H} \\ \dfrac{R_{P1}^H}{T_{P1}^H} & \dfrac{1}{T_{P1}^H} \end{bmatrix} \quad (32)$$

and by setting the denominator of $B_M^{11}$ in Eq. 18 equal to zero as:

$$\dfrac{1}{T_{F1}^H T_{F2}^H} e^{i(h_2-h_1)d} + \dfrac{R_{P1}^H R_{F2}^H}{T_{P1}^H T_{F2}^H} e^{-i(h_2+h_1)d} = 0 \quad (33)$$

the characteristic equations can be extracted. Assuming the stacked sheets are free-standing (medium 1 and medium 3 are the same), one obtains:

$$e^{ih_2 d} \mp \left( \dfrac{\dfrac{h_1}{\omega\mu_1} - \dfrac{h_2}{\omega\mu_2} - i\sigma}{\dfrac{h_1}{\omega\mu_1} + \dfrac{h_2}{\omega\mu_2} + i\sigma} \right) = 0 \quad (34)$$

Noting that $\left(e^{ih_2 d} - 1\right)/\left(e^{ih_2 d} + 1\right) = i\tan\left(\dfrac{h_2 d}{2}\right)$, the characteristic equations of even-odd TE modes can be obtained as:

$$\dfrac{h_2}{\omega\mu_2} + i\sigma + i\tan\left(\dfrac{h_2 d}{2}\right)\dfrac{h_1}{\omega\mu_1} = 0 \quad (35)$$

$$\dfrac{h_2}{\omega\mu_2} + i\sigma + i\cot\left(\dfrac{h_2 d}{2}\right)\dfrac{h_1}{\omega\mu_1} = 0 \quad (36)$$

The procedure can be repeated for TM waves by replacing the superscripts $H$ with $V$ in (33). Thus, the characteristic equations of even-odd TM modes can be obtained as:

$$e^{ih_2 d} \mp \left( \dfrac{\dfrac{\omega\varepsilon_1}{h_1} - \dfrac{\omega\varepsilon_2}{h_2} - i\sigma}{\dfrac{\omega\varepsilon_1}{h_1} + \dfrac{\omega\varepsilon_2}{h_2} + i\sigma} \right) = 0 \quad (37)$$

Therefore:

$$i\dfrac{\omega\varepsilon_1}{h_1}\tan\left(\dfrac{h_2 d}{2}\right) + \dfrac{\omega\varepsilon_2}{h_2} + i\sigma = 0 \quad (38)$$

$$\dfrac{\omega\varepsilon_1}{h_1} + i\left(\dfrac{\omega\varepsilon_2}{h_2} + i\sigma\right)\tan\left(\dfrac{h_2 d}{2}\right) = 0 \quad (39)$$

The above equations can be compared with those of [46] attained by the transverse resonance technique. Finally, the reflection ($R$) and transmission ($T$) coefficients of the plane wave from two stacked free-standing period patches in Fig. 2(a) with $D_0=5$ μm, $g_0=0.5$ μm, $\mu_c=0.5$ eV, $\tau=0.5$ ps, and $d=100$ nm are compared with the results of the TMM approach [47, 48] in Fig. 4 and a good agreement is achieved. Once the validity of the extracted coefficients is verified, the formulation can be used to calculate the interaction of any electromagnetic source with the graphene-based stacks using the convolution integral [33]. Importantly, another degree of freedom for manipulating the optical response is provided due to the arbitrary location of the source and field points in the formulas. The research can be continued by solving the Sommerfeld integrals.

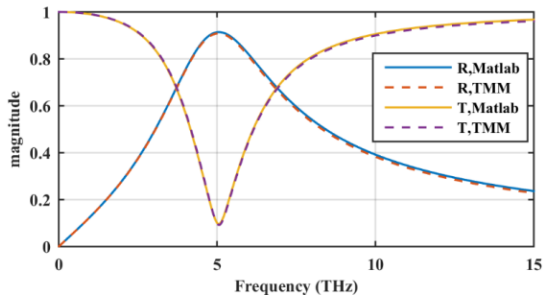

Fig 4. The reflection ($R$) and transmission ($T$) coefficients of the plane wave from two stacked free-standing period patches in Fig. 2(a) with $D_0$=5 μm, $g_0$= 0.5 μm, $\mu_c$=0.5 eV, $\tau$=0.5 ps, and $d$=100 nm.

## 4. Conclusion

Dyadic Green's function for a multi-layered planar structure with electrically biased graphene sheets at its interfaces is formulated based on the scattering superposition method. The recurrence formulas for determining the unknown coefficients of Green's function expansion are written in terms of equivalent reflection and transmission coefficients. Using dyadic formulation, the same formulation applies for TE and TM modes as long as for arbitrary sources. The extracted formulas are capable of analyzing various graphene-based planar structures such as 1D photonic crystals, filters, and metamaterials. The geometrical and material properties of the dielectric layers along with bias parameters of graphene interfaces control the optical response of the structure.

## Disclosures

The authors declare that there are no conflicts of interest related to this article